\documentstyle[12pt]{article}

\def\ha{{1\over 2}} \def\fr#1,#2{{#1\over #2}}
\def\normord#1{\mathopen{\hbox{\bf:}}#1\mathclose{\hbox{\bf:}}}
\def\ran{\rangle} \def\ket#1{|#1\ran} \def\dots{\ldots}

\begin{document}

\setcounter{page}{100}

\centerline{\large {\bf The Vacuum in the Light-Cone Representation}}
\vskip.3in \centerline{ Gary McCartor} \centerline{  SMU} \centerline{
Dallas, Texas 75275}

\vskip.6 in \centerline{\bf Abstract} \vskip.1in The mechanism by which
the physical vacuum can be different from the perturbative vacuum in
the light-cone representation is described and illustrated.  \vskip.1in

\vskip.3in {\bf 1. Introduction} \vskip.1in

In this talk I shall review the mechanism by which the physical vacuum
in  an interacting theory becomes a state other than the light-cone
perturbative vacuum.  Vacuum structure in the light-cone representation
is always associated with zero modes but there are two distinctly
different cases.  In one case the vacuum remains the physical vacuum
but some field gains a constrained zero mode due to the interaction and
that zero mode generates a nonzero, and possibly symmetry breaking,
V.E.V. for an operator which does not have one in free theory.  Such
effects have been discussed at these meetings before, especially by
Robertson$^1$, Werner et al.$^2$ and Pinsky et al.$^3$; I shall not
discuss that type of vacuum structure in this talk.  I shall discuss
the case where the interacting vacuum is a different state than the
perturbative vacuum.  That effect must occur for theories with
degenerate vacua, such as the Schwinger model, and requires the
presence of unconstrained zero modes.

I shall first review the argument that the physical vacuum is the
perturbative vacuum in the light-cone representation even for
interacting theories, and review the mechanism by which this argument
can fail.  I shall then go quickly through the examples of free theory
and the Schwinger model.  I have spoken on these cases before and the
details have been published$^{4,5}$.  I shall then apply the same
methods to the case of massless $QCD_2$.  Finally I shall speculate
briefly on $QCD_4$.

\vskip.3in {\bf 2. Vacuum Structure} \vskip.1in

The argument that the physical vacuum is the perturbative vacuum in the
light-cone representation is as follows:  the operator, $P^+$ has the
same form in an interacting theory as it does in free theory, that is,
$P^+ = P^+_{FREE}$; the physical vacuum must be an eigenstate of $P^+$
with eigenvalue 0; for theories which can be specified with
quantization conditions on $x^+ = 0$ the only such state is the
perturbative vacuum.  \vskip.1in

I want to give two arguments that $P^+ = P^+_{FREE}$ since they relate
to later things.  The first is simply to calculate the answer.  We
integrate:  $$
	      P^+ =  \ha \int T^{++} dx^-
$$ where, $$
	T^{++} = \sum_{\phi} \fr{\partial \phi},{\partial x_{+}}
	\fr{\partial {\cal L}},{\partial (\partial_{+} \phi)} - g^{++}
	{\cal L} $$ Since $g^{++}$ is zero this expression does not
depend on the interaction for nonderivative coupling.  I shall refer to
this argument as the algebraic argument.  \vskip.1in

Another argument that $P^+ = P^+_{FREE}$, in some ways more instructive
for our later work, makes use of the fact that $P^+$ is the generator
of translations within our initial value surface, $x^+ = 0$. That is:
$$
	  \partial_-\phi = {i\over 2}[P^+,\phi]
$$ Since we initialize our fields to be isomorphic to free fields on
the initial value surface, $P^+_{FREE}$ will correctly generate these
translations for all fields initialized on $x^+ = 0$.  If we have a
well posed initial value problem, and thus a complete set of fields,
the only operator we can mix with $P^+_{FREE}$ is a multiple of the
identity which would have no effect on the dynamics.  \vskip.1in

The flaw in this argument is that in the presence of massless fields
one cannot formulate a proper initial value problem with initial values
on $x^+ = 0$.  One must also specify certain zero modes---functions of
$x^+$.  Since these are true degrees of freedom they commute with the
fields specified on $x^+ = 0$ and thus can mix with $P^+$ without
contradicting the Heisenberg equations. One might ask about the
algebraic argument; I shall return to that question presently.
\vskip.1in

To be definite let us consider the case of a massless Fermi field in
$1+1$ dimensions.  We can initialize the field $\psi_+$ on $x^+ = 0$:
$$
    \psi_+(0,x^-) = {1\over\sqrt{2L}}\sum_{n=1}^\infty b(n)
   e^{-ik_-(n)x^-} + d^*(n) e^{ik_-(n)x^-} $$ The field $\psi_-$ cannot
be initialized on $x^+ = 0$ and thus furnishes the zero modes discussed
above:  $$
    \psi_-(x^+,0) = {1\over\sqrt{2L}}\sum_{n=1}^\infty \beta (n)
   e^{-ik_+(n)x^+} + \delta ^*(n) e^{ik_+(n)x^+} $$ Here we see that
any functional of $\psi_-$ could mix with $P^+$ and there would be no
contradiction with the Heisenberg equation, that is, if:  $$
	       P^+ =  P^+_{FREE} + {\cal F}(\psi_-)
$$ then still:  $$
		   \partial_-\psi_+ = {i\over 2}[P^+,\psi_+]
$$ While such mixing would not contradict this Heisenberg equation it
would contradict the full dynamics in free theory so in free theory the
$\psi_-$ modes do not mix with $P^+$.  \vskip.1in

In the Schwinger model they do.  We shall work in the gauge
$\partial_-A^+ = 0$ and find that the equations of motion are:  $$
\fr{\partial^2 A^-},{\partial x^{-2}} = -\ha {J^\prime}^+ $$ $$
-\fr{\partial^2 A^+},{\partial x^{+2}} + \fr{\partial^2 A^-},{\partial
x^+ \partial x^-} = \ha {J^\prime}^- $$ The prime on the $J$'s reflects
the need to subtract an overall zero mode from $J^0$ before coupling
the current to the Maxwell field.  We define  gauge invariant products
of Fermi fields as:  $$ \normord{\psi_+^*(x) \psi_+(x)} \equiv
 \lim_{\epsilon^-\rightarrow0} \left( e^{-ie\int_x^{x+\epsilon^-}
A_-^{(-)} dx^-} \psi_+^*(x+\epsilon^-) \psi_+(x)
e^{-ie\int_x^{x+\epsilon^-} A_-^{(+)}dx^-}-{\rm V.E.V.}\right) $$ $$
\normord{ \psi_-^*(x) \psi_-(x)} \equiv
 \lim_{\epsilon^+\rightarrow0} \left( e^{-ie\int_x^{x+\epsilon^+}
A_+^{(-)} dx^+}
 \psi_-^*(x+\epsilon^+) \psi_-(x) e^{-ie\int_x^{x+\epsilon^+} A_+^{(+)}
dx^+} - {\rm V.E.V.}\right) $$
{}From which we calculate the zero modes of the currents to be:
$$ J^{+\prime}(0) = \fr {1},{2L}Q_+ -\fr {e^2},{2\pi}A^+ - \ha  \fr
{1},{2L}Q_+ + \ha \fr {e^2},{2\pi}A^+ - \fr {1},{2L}Q_- + \ha \fr
{e^2},{2\pi}A^-(0) $$ $$ J^{-\prime}(0) = \fr {1},{2L}Q_- -\fr
{e^2},{2\pi}A^-(0) - \ha  \fr {1},{2L}Q_- + \ha \fr {e^2},{2\pi}A^-(0)
- \fr {1},{2L}Q_+ + \ha \fr {e^2},{2\pi}A^+
 = -J^{+\prime}(0) $$ Which in turn allows us to solve for the zero
modes of the gauge fields:  $$
   A^+=-{1\over Lm^2}Q_- $$ $$
   A^-(0)=-{1\over Lm^2}Q_+ $$ To find $P^+$ we must integrate the
density:  $$ T^{++} = 2i \lim_{\epsilon^-\rightarrow0} \Biggl(
 e^{-ie\int_x^{x+\epsilon^-} A_-^{(-)} dx^-} \psi_+^*(x+\epsilon^-)
\partial_- \psi_+(x) e^{-ie\int_x^{x+\epsilon^-} A_-^{(+)} dx^-}
 -  {\rm C.C.}-{\rm V.E.V.} \Biggr) $$ It is here that we see that the
algebraic argument that $P^+$ is trivial has failed.  The effect is
precisely like an anomaly:  we have a singular operator product and we
find that not all properties of the classical product can be maintained
in the quantum theory; here we must give up gauge invariance or the
purely kinematical nature of $P^+$.  For $P^+$ we get:  $$
     P^+ =  \ha \int^L_{-L} \normord{2i \Bigl(\psi^*_+ \partial_-
     \psi_+ - \partial_- \psi^*_+ \psi_+\Bigr)} dx^- = P^+_{FREE} -
{1\over 4Lm^2}Q_-^2 $$ if we define a set of special states,
$\ket{M,N}$, as:  \begin{eqnarray}
\ket{M,N}&=\delta^*\Bigl(M\Bigr)\dots \delta^*\Bigl(1\Bigr)
	d^*\Bigl(N\Bigr)\dots d^*\Bigl(1\Bigr)\ket0 \qquad(M>0,N>0)
\nonumber \\ \nonumber \ket{M,N}&=\beta^*\Bigl(M\Bigr)\dots
\beta^*\Bigl(1\Bigr)
	d^*\Bigl(N\Bigr)\dots d^*\Bigl(1\Bigr)\ket0
\qquad(M<0,N>0)\nonumber \\ \ket{M,N}&=\delta^*\Bigl(M\Bigr)\dots
\delta^*\Bigl(1\Bigr)
	b^*\Bigl(N\Bigr)\dots b^*\Bigl(1\Bigr)\ket0
\qquad(M>0,N<0)\nonumber \\ \nonumber
\ket{M,N}&=\beta^*\Bigl(M\Bigr)\dots \beta^*\Bigl(1\Bigr)
	b^*\Bigl(N\Bigr)\dots b^*\Bigl(1\Bigr)\ket0 \qquad(M<0,N<0)
\end{eqnarray} we find that:  $$
      P^+\ket{M,N} = 0 $$ For $M = -N$ these states are in the physical
subspace and form the degenerate ground states of the Schwinger model.
To form a $\theta$-state we take:  $$
		\ket{\theta} = \sum e^{iM\theta} \ket{M,-M}
$$ \vskip.1in

The point is not just that degenerate ground states can be accommodated
within the light-cone representation but that there is a limited number
of ways that that can occur.  That fact leads, in the case of the
Schwinger model, to the fact that the ground states are much simpler in
the light-cone representation than they are in the equal-time
representation---simpler to express and simpler to find.  It also
suggests that the light-cone representation may be useful in the
analysis of vacuum structure for more complicated theories; the number
of operators which can mix with $P^+$ grows rather slowly with the
dimension of space-time and the vacuum structure is controlled
substantially by $P^+$.  \vskip.1in

Let us now apply these same considerations to $QCD_2$ with color group
$SU(2)$ and quarks in the fundamental representation.  We initialize
the fields as before:  $$
    \psi_+^i(0,x^-) = {1\over\sqrt{2L}}\sum_{n=1}^\infty b^i(n)
   e^{-ik_-(n)x^-} + d^{i*}(n) e^{ik_-(n)x^-} $$ $$
    \psi_-^i(x^+,0) = {1\over\sqrt{2L}}\sum_{n=1}^\infty \beta^i (n)
   e^{-ik_+(n)x^+} +    \delta^{i*}(n) e^{ik_+(n)x^+} $$ where the
$i$'s are color indices.  Gauge invariant currents are given by, for
instance:  $$ \normord{\psi_+^*(x)T^a \psi_+(x)} \equiv
 \lim_{\epsilon^-\rightarrow0} \left( e^{-ie\int_x^{x+\epsilon^-}
A_-^{b(-)}T^b_{jk} dx^-} \psi_+^{i*}(x+\epsilon^-)T^a_{ij} \psi_+^k(x)
e^{-ie\int_x^{x+\epsilon^-} A_-^{c(+)}T^c_{jk} dx^-}-{\rm
V.E.V.}\right) $$ We find for the currents:  $$
       J^{a+}(0,x^-) = \fr {g},{2L}\sum_{n=1}^\infty \Biggl(C^{a*}(n)
      e^{ik_-(n)x^-}- C^a(n)e^{-ik_-(n)x^-}\Biggr) + {g\over{2
   L}}Q_+^a  - {g^2\over 4\pi}A^{a+} $$ $$
       J^{a-}(x^+,0) = \fr {g},{2L}\sum_{n=1}^\infty \Biggl(D^{a*}(n)
      e^{ik_+(n)x^+}- D^a(n)e^{-ik_+(n)x^+}\Biggr) + {g\over{2
   L}}Q_-^a  - {g^2\over 4\pi}A^{a-} $$ where the $C$'s and $D$'s are
the fusion operators and the $Q$'s are the charges.  A fully
symmetrized equation of motion for the gauge field is:
\begin{eqnarray} -\partial_-^2A^{1-} &+ \fr{g},{2} \Bigl(
-A^{2+}\partial_-A^{3-} - \partial_-A^{3-}A^{2+} +
A^{3+}\partial_-A^{2-} + \partial_-A^{2-}A^{3+}\nonumber \\ & +\ha
A^{2+}\partial_+ A^{3+} +\ha \partial_+ A^{3+}A^{2+} -\ha
A^{3+}\partial_+A^{2+} -\ha \partial_+A^{2+}A^{3+} \Bigr)\nonumber
\\ \nonumber &+\fr{g^2},{16} \Bigl( A^{1-}A^{2+}A^{2+} +
A^{2+}A^{2+}A^{1-} + 2 A^{2+}A^{1-}A^{2+}\nonumber \\ &-
A^{2-}A^{1+}A^{2+} - A^{2+}A^{1+}A^{2-} - A^{1+}A^{2-}A^{2+} -
A^{2+}A^{2-}A^{1+}\nonumber \\ &- A^{3-}A^{1+}A^{3+} -
A^{3+}A^{1+}A^{3-} - A^{1+}A^{3-}A^{3+} - A^{3+}A^{3-}A^{1+}\nonumber
\\ \nonumber &+ A^{1-}A^{3+}A^{3+} + A^{3+}A^{3+}A^{1-} + 2
A^{3+}A^{1-}A^{3+} \Bigr)  = \ha {J^\prime}^{1+} \end{eqnarray} and
similarly for the other Maxwell equation.  The zero modes of these
equations are:  \begin{eqnarray}
 &\fr{g},{4} ( A^{2+}\partial_+ A^{3+} + \partial_+ A^{3+}A^{2+} -
 A^{3+}\partial_+A^{2+} - \partial_+A^{2+}A^{3+} )\nonumber  \\
&+\fr{g^2},{16} ( A^{1-}A^{2+}A^{2+} + A^{2+}A^{2+}A^{1-} + 2
A^{2+}A^{1-}A^{2+}\nonumber \\ &- A^{2-}A^{1+}A^{2+} -
A^{2+}A^{1+}A^{2-} - A^{1+}A^{2-}A^{2+} - A^{2+}A^{2-}A^{1+}\nonumber
\\ &- A^{3-}A^{1+}A^{3+} - A^{3+}A^{1+}A^{3-} - A^{1+}A^{3-}A^{3+} -
A^{3+}A^{3-}A^{1+}\nonumber \\ &+ A^{1-}A^{3+}A^{3+} +
A^{3+}A^{3+}A^{1-} + 2 A^{3+}A^{1-}A^{3+} )\nonumber \\ \nonumber &  =
-\ha \Bigl( - {g\over{2 L}}Q_-^a  - {g^2\over 4\pi}A^{a+} \Bigr)
\end{eqnarray} \begin{eqnarray} &-\partial_+^2A^{1+} + \fr{g},{2}
\Bigl( -A^{2-}\partial_+A^{3+} - \partial_+A^{3+}A^{2-} +
A^{3-}\partial_+A^{2+} + \partial_+A^{2+}A^{3-}  \Bigr) \nonumber \\
&+\fr{g^2},{16} \Bigl( A^{1+}A^{2-}A^{2-} + A^{2-}A^{2-}A^{1+} + 2
A^{2-}A^{1+}A^{2-}\nonumber \\ &- A^{2+}A^{1-}A^{2-} -
A^{2-}A^{1-}A^{2+} - A^{1-}A^{2+}A^{2-} - A^{2-}A^{2+}A^{1-}\nonumber
\\ &- A^{3+}A^{1-}A^{3-} - A^{3-}A^{1-}A^{3+} - A^{1-}A^{3+}A^{3-} -
A^{3-}A^{3+}A^{1-}\nonumber \\ \nonumber &+ A^{1+}A^{3-}A^{3-} +
A^{3-}A^{3-}A^{1+} + 2 A^{3-}A^{1+}A^{3-} \Bigr)  = -\ha  \Bigl( -
{g\over{2 L}}Q_+^a  - {g^2\over 4\pi}A^{a-} \Bigr) \end{eqnarray} I do
not think these equations can be implemented at the operator level.  A
construction which works is as follows:  Set $$
   A^{a+}=-{2\pi\over Lg}Q_-^a \quad;\quad A^{a-}(0)=-{2\pi\over
   Lg}Q_+^a $$ $$
	P^+ = P^+_{FREE} - {\pi\over 2L} (Q_-^aQ_-^a) $$ $$
	 P^- = P^-_{FREE} - {\pi\over 2L} (Q_+^aQ_+^a) + {g^2\over 2}
	 \sum {1 \over k_-^2(n)} C^{a*}(n)C^a(n)
$$ Define the physical subspace by:  $$
		 D^a(n)\ket{{\cal P}} = 0 \qquad a = 1,2,3
$$ $$
	 (Q_-^aQ_-^a)\ket{{\cal P}} = (Q_+^aQ_+^a)\ket{{\cal P}} = 0
$$ The unexpected thing is that these last two restrictions hold
separately.  It is that fact that holds the ground state of the system
in the perturbative vacuum.  If the restriction were the naively
expected:  $$
     (Q_-^aQ_-^a + Q_+^aQ_+^a)\ket{{\cal P}} = 0 $$ the perturbative
vacuum would decay into some combination of $\ket{1} =
(d^{1*}(1)\beta^{1*}(1) + d^{2*}(1)\beta^{2*}(1))\ket{0}$ and $\ket{2}
= (\delta^{1*}(1)b^{1*}(1) + \delta^{2*}(1)b^{2*}(1))\ket{0}$.  The
physical effect is that long range interactions ( the zero modes )
stabilize the the perturbative light-cone vacuum and prevent the
occurrence of degenerate ground states.  Eric Zhitnitsky has told me
that Andi Smilga has reached a similar conclusion on the basis of
lattice calculations.

\vskip.3in {\bf 3. Further Work} \vskip.1in

Issues in $1+1$ dimensions which would be interesting to examine
include the problem of adjoint matter, which semiclassical arguments
suggest should be different from fundamental matter, and whether or not
twisted boundary conditions make a difference$^6$.  In four dimensions
one expects the operators which can mix with $P^+$ to be formed from
sixteen gluon and six quark fields, in each case a function of the
single variable $x^+$.  The problem is to calculate the mixing.  As is
seen from the discussion above, operator mixing induced by
renormalization plays a central role in the analysis.  In higher
dimensions that problem is more important and more difficult.  Indeed
that problem is central not only to the zero mode problem but the whole
field of light-cone techniques---as the Ohio State group keeps
reminding us.  While the problem is difficult and I do not know how to
solve it yet, I believe the light-cone representation may prove to be a
valuable tool in the analysis of vacuum structure.

\vskip.3in {\bf Acknowledgement} \vskip.1in

I thank the organizers for their kind invitation to speak here and for
their warm hospatility during the meeting.

\vskip.3in {\bf References} \vskip.1in

1. D.G. Robertson, Phys. Rev. {\bf D47}, 2549 (1993).

2. T. Heinzl, S. Krusche, S. Simb\"urger, and E. Werner, Z. Phys. {\bf
C56}, 415 (1992).

3. S. Pinsky and B. van de Sande, Phys. Rev. {\bf D49}, 2001 (1994).

4. G. McCartor, Z. Phys. {\bf C52}, 611 (1991)

5. G. McCartor, {\it Schwinger Model in the Light-Cone Representation},
Z. Phys. {\bf C}, {\it in press}.

6. M. A. Shifman and A. V. Smilga, {\it Fractons in Twisted Multiflavor
Schwinger Model}, UNM-TH-1262-94, July 1994

\end{document}